# Ferromagnetism and magneto-dielectric effect in insulating LaBiMn$_{4/3}$Co$_{2/3}$O$_6$ thin films

R. Ranjith, Asish K. Kundu, M.Filippi, B.Kundys, W. Prellier[*], B. Raveau

Laboratoire CRISMAT, CNRS UMR 6508, ENSICAEN,

6 Boulevard Maréchal Juin, Caen-14050, France

J. Laverdière, M.P. Singh and S. Jandl

Department de physique & Regroupement québécois sur les materiaux de pointe, Université de Sherbrooke, Sherbrooke, Québec J1K2R1, Canada.

High quality epitaxial thin films of LaBiMn$_{4/3}$Co$_{2/3}$O$_6$ perovskite were fabricated on (001)-oriented SrTiO$_3$ and LaAlO$_3$ substrates by the pulsed laser deposition technique. Magnetization measurements reveal a strong magnetic anisotropy and a ferromagnetic behavior that is in agreement with a super-exchange interaction between Mn$^{4+}$ and Co$^{2+}$ ions, which are randomly distributed in the B-site. A distinct anomaly is observed in the dielectric measurements at 130K corresponding to the onset of the magnetic ordering, suggesting a coupling. Above this temperature, the extrinsic Maxwell-Wagner effect is dominating. Theses results are explained using the Raman spectroscopic studies indicating a weak spin-lattice interaction around this magnetic transition.

---

[*] Corresponding author: prellier@ensicaen.fr

Multifunctional materials (multiferroics, magneto dielectric, spintronics etc.) are of interest owing to their potential device applications, such as non-volatile memories, magnetic read heads, tunnel junction spin filtering etc[1]. Materials that exhibit such behavior possess a coupling between electronic and magnetic properties, which is reflected in their respective order parameters. Magnetic semiconductors or insulators have been identified as single phase materials that offer the potential to exhibit simultaneously electronic and magnetic ordering, and until now, very few compounds of this type have been reported (e.g. $BiMnO_3$,[2,3] $CdCr_2Se_4$,[4] $La_2Mn(Co/Ni)O_6$,[5,6] $Bi_2MnNiO_6$,[7,8] and diluted magnetic semiconductors [9]). In the metal oxide members containing bismuth, ferromagnetism originates from super-exchange interactions between adjacent cations through the oxygen and ferroelectricity likely originates from the lone pair electrons of the $Bi^{3+}$ ion, which induce structural distortions. As a result, the investigation of ferromagnetic insulators containing bismuth offers the potential to generate magnetoelectric properties. In this respect, the insulating $LaBiMn_{4/3}Co_{2/3}O_6$ (LBMCO) perovskite is attractive since at low temperature it is a hard ferromagnet (useful for memory device).[10] This material has a higher ferromagnetic Curie temperature (130K) than $BiMnO_3$ ($T_C\sim105K$) and can be prepared at normal pressure. LBMCO possesses a high value of thermoelectric power at room temperature and p-type polaronic conductivity. Nevertheless, no detailed ferroelectric studies have been reported for this compound. However, recent investigations of the magneto-dielectric[11] or spin filtering,[12] for these Bi-based materials in the form of thin films have enhanced the possibility of device applications.

LBMCO films (1500Å) were deposited on (001)-oriented $SrTiO_3$ (STO) and $LaAlO_3$ (LAO) substrates supplied by CrysTec (GmbH, Germany) by the pulsed laser deposition (PLD). A KrF excimer laser ($\lambda$ = 248nm, 3Hz) was focused on the dense ceramic target of LBMCO, prepared through conventional sol-gel method.[10] The deposition was carried out at a substrate temperature of 650°C under 20 mTorr of oxygen. The crystalline quality and the epitaxial nature of the films were confirmed by x-ray diffraction ($\lambda$=1.5406Å) using normal θ-2θ configuration (Seifert) and Φ-scans (Phillips Xpert). The cationic ordering and the spin-lattice interactions were investigated by polarized Raman spectroscopy using a Labram-800 microscope spectrometer equipped with a He:Ne laser($\lambda$=632.8nm) and nitrogen cooled CCD detector. The magnetization (M) and the capacitance (C) measurements were investigated as a function of temperature (T) and magnetic field (H) by a Quantum Design AC SQUID and physical property measurement system (PPMS) coupled with an impedance analyzer (Agilent technologies- 4284A), respectively.

Figure 1 shows the Θ−2Θ x-ray diffraction pattern of the LBMCO thin film. The diffraction pattern reveals that the films are highly oriented and no alternate phases or extra orientations were detected. Bulk LBMCO crystallizes in a *Pnma* space group with lattice parameters of $a_{bulk}$=5.529 Å, $b_{bulk}$=7.801 Å and $c_{bulk}$=5.513 Å.[10] Consequently, the lattice parameters of the film would have the following the relationships: $a_P$=$c_P$=$a_{bulk}/\sqrt{2}$=3.905 Å, b=2$a_P$,[13] where $a_P$ refers to the lattice parameter of the cubic perovskite sub-cell (close to 3.9Å). Based on this, the films would have either a [010] or a [101]-orientation with respect to the substrate plane. Such an orientation does not affect the properties and the micro-structural characterizations will be published elsewhere. On LAO, the out-of-plane lattice parameter is calculated to be 3.907Å. Such a value matches well with STO substrate (3.905 Å), as can be seen from the overlap of the diffraction peaks. The out-of-plane lattice parameter is slightly different than the bulk lattice parameter by +0.2%, when assuming as a [010]-orientation, and -0.08%, if a [101]-orientation of the film is assumed. The Φ-scan recorded around (103) reflection of the film is shown in the inset of Figure 1. The presence of the 4 peaks, 90°-separated confirms the cube-on-cube growth of the thin film with respect to the substrate, indicating the epitaxial nature of the film.

Figure 2 shows the temperature dependent zero-field-cooled (ZFC) and field-cooled (FC) magnetization of the LBMCO film on LAO substrate. Magnetic field was applied both parallel (H||(100)$_S$) and perpendicular (H||(001)$_S$) to the substrate surface. At 10K, thin films exhibited a higher magnetization (2.21$\mu_B$/f.u (FC)) for the parallel applied magnetic field parallel to substrate (film) surface (H||(100)$_S$) in comparison to the perpendicular applied magnetic field (H||(001)$_S$) (0.68$\mu_B$/f.u.(FC)). A strong magnetic anisotropy (along the directions parallel and perpendicular of the film plane) can be associated with a single domain or epitaxial orientation of the LBMCO films (on both STO and LAO substrates). The inset of Figure 2 shows the magnetic hysteresis loop, (M-H), for the LBMCO films recorded at 10 and 100K. At 10K, a remnant magnetization ($M_r$) value of 2.26 $\mu_B$/f.u. and 2.43 $\mu_B$/f.u. and a coercive field ($H_C$) value of 8.0 and 7.1 kOe are observed for LBMCO films deposited on LAO and STO, respectively. As expected, the high temperature (M-H) behavior is linear similar to paramagnetic phase and dominated by the diamagnetic behavior of the substrate. The obtained (M-H) plots and the strong divergence between ZFC-FC magnetization at low temperature are similar to those of the bulk material reported earlier. On the other hand, the $H_C$ values for the LBMCO thin films are greater than those of the bulk phase.[10] The small variation in the $H_C$ as well as in the Curie temperature ($T_C$) values observed in LBMCO films on different substrates may be due to the tensile strain of the films. The ZFC and FC magnetization data of LBMCO films exhibit a clear transition from a paramagnetic state to a ferromagnetic state at $T_C$. The $T_C$ values were calculated from the minimum position of the $dM_{FC}/dT$ versus temperature. A $T_C$ value of ~123K and 115K is obtained for the film deposited on LAO and STO substrates, respectively, which is in close agreement with the bulk(130K).[10] A large divergence between ZFC and FC data at low temperature was observed for both substrates. The $T_{irr}$ (the temperature at which, ZFC and FC magnetization diverges) is also higher for LAO film ($T_{irr}$ ~100K) whereas for STO film the $T_{irr}$ value is 85K. Aforementioned, low temperature magnetic properties prove the existence of ferromagnetism in LBMCO film, and are in close agreement with the possibility of FM interaction between $Mn^{4+}$ and $Co^{2+}$ cations via super-exchange mechanism.[4-6] The spin-glass behavior of the low temperature magnetic phase has been reported previously in the bulk.[10]

Figure 3 shows the dielectric measurements carried out at 1 MHz in the temperature range of 10 to 300K, both for bulk and thin film samples. In contrast to the film, the bulk exhibits a huge rise in the dielectric constant above 150K owing to an extrinsic Maxwell-Wagner kind relaxation, which arises from the grain boundaries present in the system[14]. Above 250K, the dielectric constant further increases in the bulk samples, which can be attributed to the space charge effect with an increase in temperature, owing to the semiconducting nature of the system. Around 200K, the curve of dielectric constant, measured at 1MHz, exhibits an anomaly in both the thin film and the bulk. These features are frequency dependent (not shown) as expected from extrinsic effect.[14]. Another dielectric anomaly is observed around the magnetic transition temperature (123K), for thin film samples. This anomaly argues for the presence of a weak spin lattice interaction in the thin films around the magnetic transition. The onset of interactive ordered magnetic clusters could give rise to spin lattice coupling, which distorts the lattice, and alters the dielectric constant[15]. The presence of lone pair electrons of bismuth can facilitate this lattice distortion at the onset of magnetic ordering. Nevertheless, the specific details of the origin and the nature of the observed weak spin-lattice interaction remain ambiguous, and are currently under study in the case of both bulk and thin films.

The dielectric anomaly around the magnetic transition observed in thin film samples is dominated by the extrinsic Maxwell-Wagner effect at low frequencies (< 500kHz). The imaginary dielectric constant ($\varepsilon''$) was observed to be in the range of 245 – 150 within the observed temperatures of 10-300K respectively. The inset of Figure 3 shows the isothermal magnetic field-dependence of the dielectric constant measured on thin film at 10K, under 1MHz. A positive magnetodielectric effect $\Delta\varepsilon$ (where $\Delta\varepsilon$ is calculated using the following formula $[(\varepsilon(H)- \varepsilon(0))/\varepsilon(0)]\times 100$ %) of 0.7% was observed at 10K. This small effect was observed up to 100K, which is close to the magnetic transition temperature of 123K, without the overlap of other extrinsic effects. The weak magnetodielectric effect (+0.7%) observed could arise from the intrinsic spin-lattice coupling.[16]

To further investigate the presence of weak spin – lattice coupling, the local structure of the thin films was investigated by polarized Raman spectroscopy.[17,18] Temperature dependence Raman spectra were measured on film samples grown on both STO and LAO surfaces in the x'x', xx, xy, and x'y' polarization configurations. A typical Raman spectra collected at 300K is shown in Figure 4. In orthorhombic perovskite crystal symmetry, 24 Raman modes are allowed, but only 4 Raman modes were observed in these samples. These modes appear at 270 cm$^{-1}$ (xx and x'x' configurations), 518 cm$^{-1}$ (not seen in x'y'), 626 cm$^{-1}$ (xy and x'x') and 635 cm$^{-1}$ (xx and x'x'). Detailed studies further show that the Raman peaks are characterized by a very large full-width at half maxima (FWHM) value. For example, the FWHM value for the 635 cm$^{-1}$ phonon mode ($A_g$-mode) is about 60 cm$^{-1}$. Such broadness in the phonon peaks indicates that the cations are randomly distributed, leading to large cationic disorder among the Mn$^{4+}$ and Co$^{2+}$. Moreover, the apparent lack of polarization dependence further attests to the lack of a cation ordering. To observe the spin-lattice interactions, a temperature dependent shift of the 633 cm$^{-1}$ Raman mode was plotted for films grown on STO and LAO substrates (see inset of Fig. 4). This plots reveals that as the temperature decreases, the shift in frequency first peaks up around 230K for STO and 250 K for LAO, respectively, and further decreases until 80K, after which it stays constant. The observed softening trend is consistent with what has been reported for other double perovskite manganites. The modulation of the super-exchange integral by the phonons is generally observed as softening of phonons[19], which occurs in the vicinity of the magnetic transition temperature. In the present case, however, the magnetic transition temperature is about 123K for bulk (see Fig. 2), which is much lower than the observed phonon peak shifts at 230 K (see Fig. 4). It is also important to note that this temperature behavior correlates well with the temperature dependence dielectric properties of these films, which also peak around the same temperature. The same ambiguity has been also observed in other magnetodielectric materials, such as DyMnO$_3$[20].

In summary, highly oriented epitaxial LaBiMn$_{4/3}$Co$_{2/3}$O$_6$ thin films were grown. The films deposited, on STO and LAO substrates, exhibited a clear ferromagnetic behavior with a Curie temperature of 123K and 115K, respectively, and a saturation magnetization of 2.21$\mu_B$/f.u. for the applied magnetic field parallel to the film surface and 0.68$\mu_B$/f.u. for the applied magnetic field perpendicular to the film surface. The high frequency anomaly observed in the dielectric measurements of the thin films around the onset of magnetic ordering temperature, suggests the presence of a weak spin-lattice interaction. A weak magneto-dielectric effect of 0.7% at 10K arises from the existing spin-lattice interaction was also confirmed by observation of softening of phonon modes in polarized Raman spectroscopic studies.

AKK thanks the French Ministry of Education and Research for a fellowship award. This work was carried out in the frame of the NoE FAME (FP6-5001159-1), the STREP MaCoMuFi (NMP3-CT-2006-033221), and the STREP CoMePhS (NMP4-CT-2005-517039) supported by the European Community and by the CNRS, France. Partial support from the ANR (NT05-1-45177, NT05-3-41793) is also acknowledged. The authors would also like to acknowledge Dr. L. Mechin, Mr. C. Fur for their help in the experiments and W.C. Sheets, M. Maglione and J.F. Scott for careful reading of the manuscript.

**Figure captions**

**Figure 1.** XRD pattern of LBMCO thin films on STO and LAO substrates. Inset shows the Φ-scan around (103) reflection of the thin film.

**Figure 2.** (color online) Temperature dependent ZFC (open symbol) and FC (solid symbol) magnetization (M), of LaBiMn$_{4/3}$Co$_{2/3}$O$_6$ thin film on LaAlO$_3$ (001) substrates (H = 1000 Oe, applied parallel to film surface H∥(100)$_S$, and perpendicular to the film surface, H∥(001)$_S$. The insets show the magnetic hysteresis curves at different temperatures (H∥(100)$_S$). Contribution from the substrate was substracted.

**Figure 3.** (color online) Variation of the relative permittivity ($\varepsilon_r$) with temperature for LBMCO compound (left for bulk, and right for thin film). The inset shows the variation of Δε % {[ε (H)- ε (0)/ ε (0)] x 100} with magnetic field at 10K for LBMCO/LAO film.

**Fig. 4.** (color online) Raman spectrum of the LBMCO/LAO in both HH and HV geometry. Inset shows the softening of a Raman mode with temperature in both substrates.

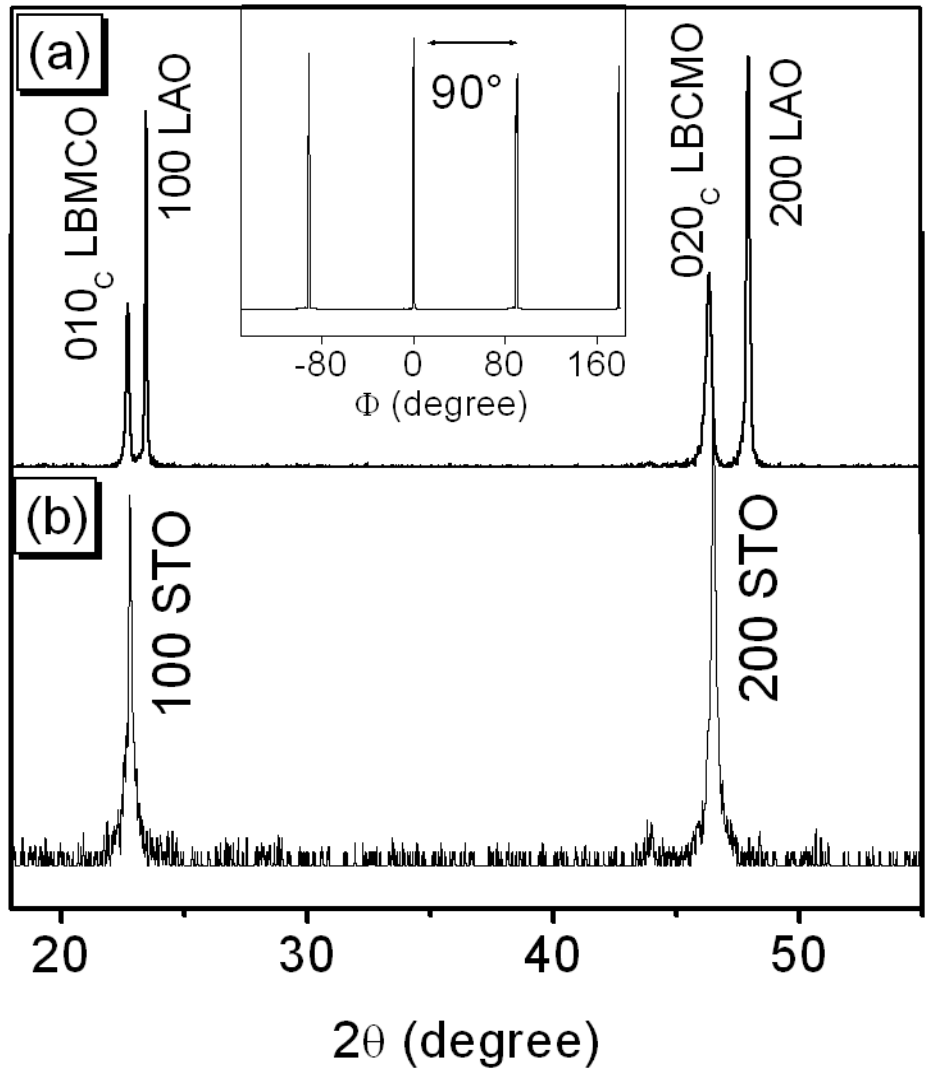

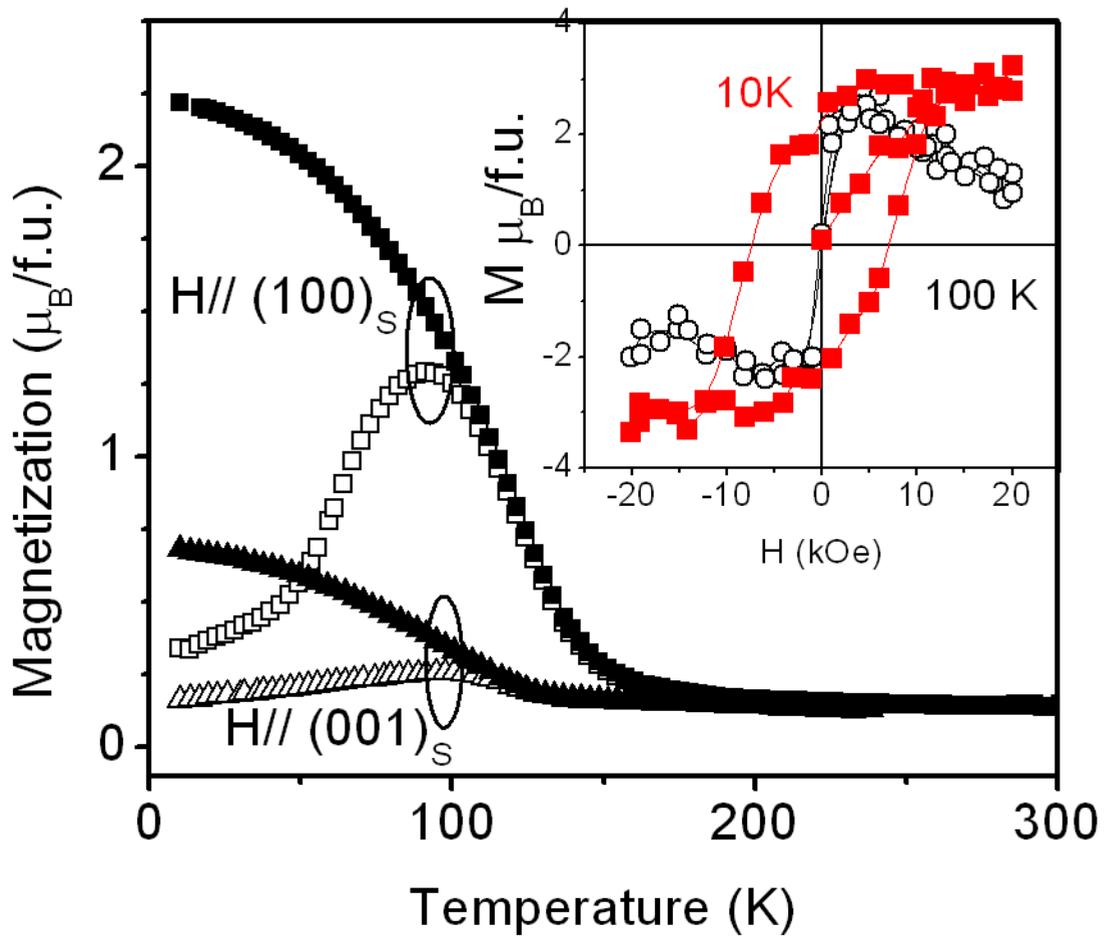

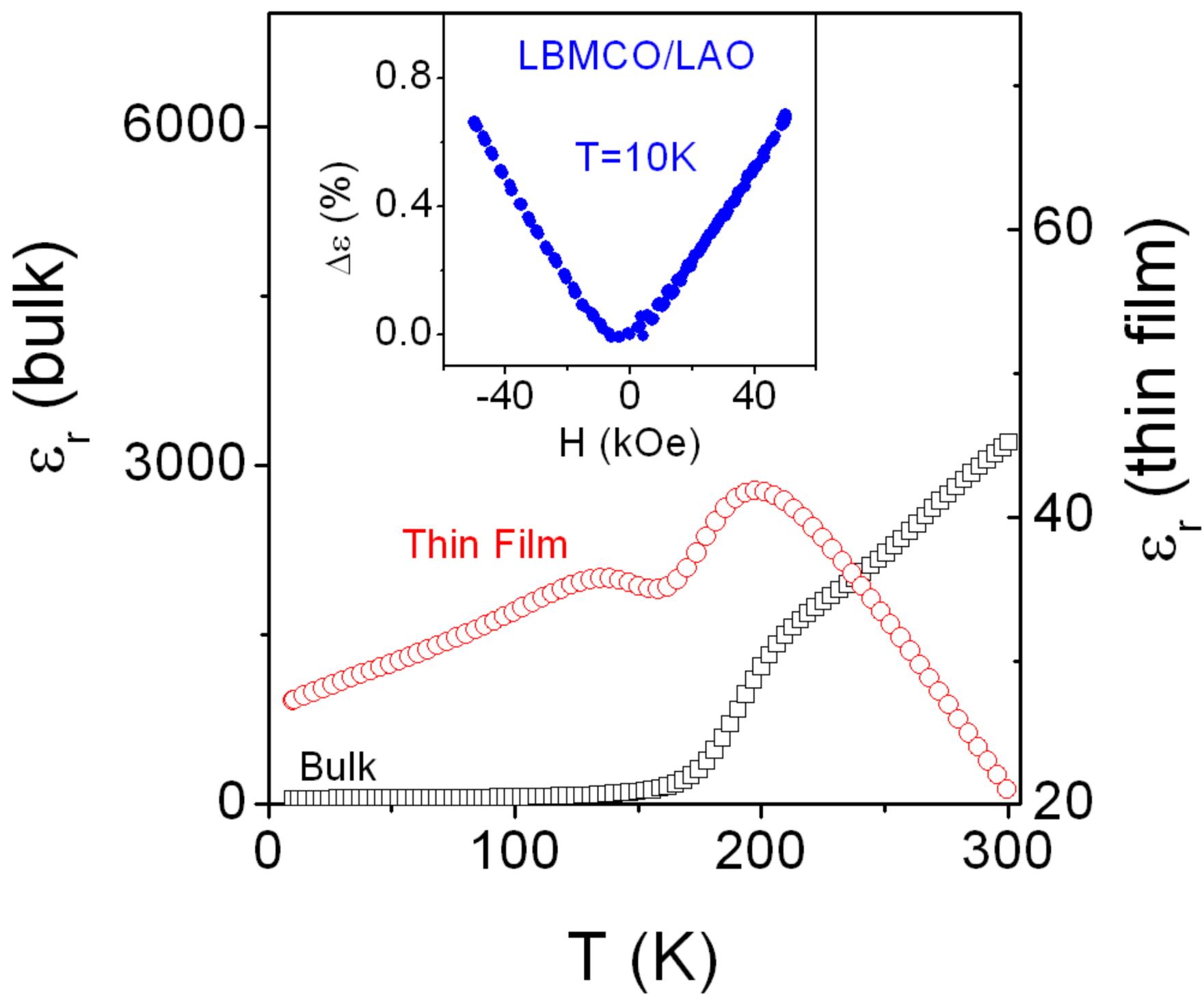

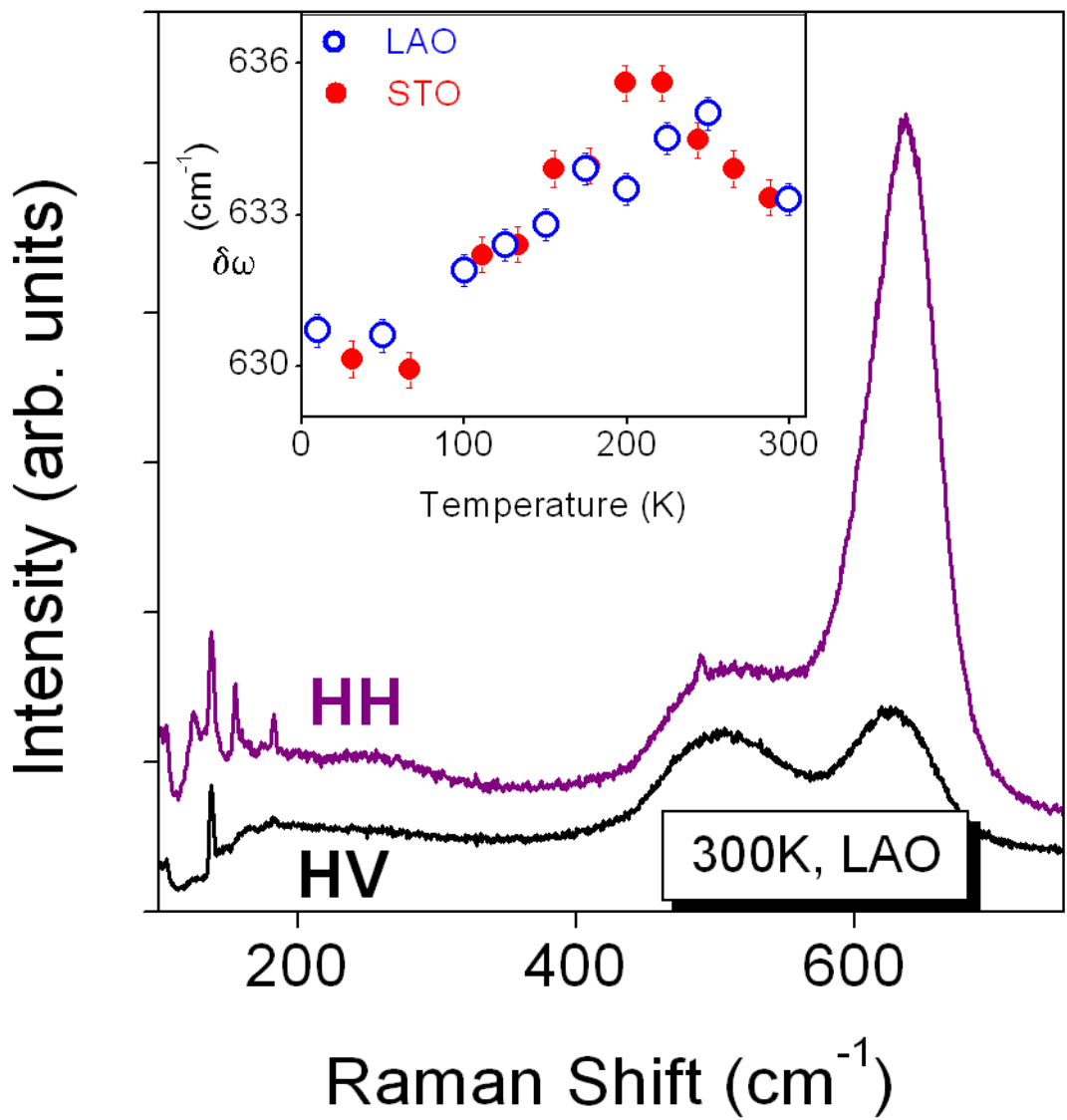